\documentclass[fleqn,12pt]{article}
\usepackage{amssymb,graphics,epsfig,palatino}

\newcommand{\bequ}{\begin{equation}}
\newcommand{\eequ}{\end{equation}}
\newcommand{\beqd}{\begin{displaymath}}
\newcommand{\eeqd}{\end{displaymath}}
\newcommand{\beqn}{\begin{eqnarray}}
\newcommand{\eeqn}{\end{eqnarray}}
\newcommand{\bsl}{\begin{sloppypar}}
\newcommand{\esl}{\end{sloppypar}}

\begin{document}

\pagestyle{empty}

\vspace*{-1cm}
\begin{flushright}
  CERN-PH-TH/2005-014\\
  DCPT-04-114\\ 
  IPPP-04-57\\  
  TSL/ISV-2005-0287\\
  UWThPh-2004-28 \\
  WUE-ITP-2005-001\\
  hep-ph/0502036
\end{flushright}

\vspace*{1.4cm}

\begin{center}

{\Large {\bf 
Distinguishing between MSSM and NMSSM\\ by combined LHC and ILC
analyses
}}\\

\vspace{10mm}

{\large
G.~Moortgat-Pick$^{a}$,
S.~Hesselbach$^{b,c}$,
F.~Franke$^d$, H.~Fraas$^d$ 
}

\vspace{6mm}

$^a${\it TH Division, Physics Department, CERN, CH-1211 Geneva 23, 
Switzerland}\\
$^b${\it Institut f\"ur Theoretische Physik, Universit\"at Wien, A-1090
Vienna, Austria}\\
$^c${\it High Energy Physics, Uppsala University, Box 535, S-751 21 Uppsala,
  Sweden}\\
$^d${\it Institut f\"ur Theoretische Physik und Astrophysik, 
Universit\"at W\"urzburg,\\
Am Hubland,}
{\it D-97074 W\"urzburg, Germany}\\

\end{center}

\vspace{1cm}

\begin{abstract}
We show that the interplay between the LHC and the $e^+ e^-$ International
Linear Collider (ILC) with $\sqrt{s}=500$~GeV might be crucial for the
discrimination between the minimal and next-to-minimal supersymmetric
standard model. We present an NMSSM scenario that cannot be distinguished
from the MSSM by cross sections and mass measurements if only the light
neutralinos and the lightest chargino are kinematically accessible, even
if one of the neutralinos has a significant singlino component.
Mass predictions for the heavier neutralinos from the ILC analysis 
and their observation at the LHC lead to an identification of the
neutralino mixing character and the underlying supersymmetric model
in a combined LHC/ILC analysis.
In our numerical example we include errors in the mass
measurements and use standard methods of supersymmetric parameter
determination.
\end{abstract}

\newpage
\pagestyle{plain}
\section{Introduction}
\label{sec:intro}
The International Linear Collider (ILC) is intended to start with an energy of $\sqrt{s}=500$~GeV,
which will be upgraded to about 1~TeV 
\cite{ITRP}. 
The physics program includes the
complement of the results at the 
Large Hadron Collider (LHC) \cite{Branson:2001ak}
and the discovery of possible signatures of physics beyond the standard
model (SM). Since supersymmetry is often considered as the main
candidate for new physics, the precise resolution of the 
supersymmetric model is one of the essential goals 
of the ILC 
\cite{TDR}. 

An interesting
possibility for the determination of the supersymmetric model 
is to study the
gaugino/higgsino particles, which are expected to be among the lightest
supersymmetric particles. In this paper we consider 
two basic supersymmetric models:
the minimal supersymmetric standard model (MSSM) and the
next-to-minimal supersymmetric standard model (NMSSM).
The MSSM contains four
neutralinos $\tilde{\chi}^0_i$, 
which are mixtures
of the photino, zino and neutral
higgsinos, and two charginos $\tilde{\chi}^\pm_i$, 
being mixtures of wino and charged 
higgsino. The neutralino/chargino sector depends at tree level
on four parameters: 
the U(1) and SU(2) gaugino masses
$M_1$ and $M_2$, 
the higgsino mass parameter $\mu$, and 
the ratio 
$\tan\beta$
of the vacuum expectation values of the Higgs fields.
For the determination of these parameters, straightforward
strategies 
\cite{parameters,choi} 
have been worked out
even if only the
light neutralinos and charginos $\tilde{\chi}^0_1$,
$\tilde{\chi}^0_2$ and $\tilde{\chi}^\pm_1$ are kinematically
accessible at the first stage of the ILC~\cite{ckmz}.   

The NMSSM \cite{nmssm} 
is the simplest extension of the MSSM 
by an additional Higgs singlet field.  
New parameters in the neutralino sector are the vacuum expectation
value $x$ of the singlet field and
the trilinear couplings $\lambda$ and $\kappa$ in the superpotential,
where the product $\lambda x = \mu_{\rm eff}$ replaces the
$\mu$-parameter of the MSSM \cite{Franke,Choi:2004zx}.
The additional fifth neutralino may significantly change the phenomenology 
of the neutralino sector.
In scenarios where the lightest supersymmetric particle is a nearly
pure singlino, the existence of displaced vertices leads
to a particularly interesting experimental signature
\cite{singlinohugonie,Singlinos}. 
If the complete neutralino sector is accessible, the 
sum rules
for the production cross sections 
show a different energy dependence in MSSM and NMSSM \cite{ckmz}. 
The NMSSM Higgs sector also depends on the soft scalar mass parameters
$A_\lambda$ and $A_\kappa$. It contains five physical neutral
Higgs bosons, three scalars $S_i$ and two pseudoscalars $P_i$.
In the following we will assume
that only the light neutralinos, charginos and 
the lightest Higgs boson $S_1$ are
accessible at the first stage of the linear collider. 

It has already been worked out that there exist
MSSM and NMSSM scenarios with
the same mass spectra of the light neutralinos but
different neutralino mixing. In this case
beam polarization is crucial for distinguishing the two models
\cite{Sitges}.
We present a scenario where
all kinematically accessible neutralinos and charginos have
similar masses and almost identical cross sections, 
within experimental errors, in MSSM and NMSSM.
Although the second lightest neutralino in the NMSSM has a significant 
singlino component, 
the models cannot be distinguished by precision measurements of
masses and cross sections at the LHC or at the ILC with $\sqrt{s}=500$
GeV. While the branching ratios for the neutralino decays into
Higgs bosons may give first evidence, final identification of the
underlying model requires precision measurements of the masses and
mixing characters of the heavier neutralinos by combined
analyses of LHC and ILC$_{500}$ results and by 
exploiting the ILC$_{650}^{{\cal L}=1/3}$ option with 
$\sqrt{s}=650$~GeV at the cost of reduced luminosity. 
The ILC$_{650}$ option 
reflects the high flexibility of the ILC and does not
require any specific setups in the experiment.
Of course, the second energy stage of the ILC with energies 
up to 1 TeV also allows a discrimination.
Here, however, we focus on the potential of the ILC$_{500/650}$.

This paper is organized as follows: 
in Section 2 we describe in detail our strategy for the discrimination
between NMSSM and MSSM. 
Our analysis is based on the scenario presented in Section 3.
The following sections 4, 5 and 6 contain the numerical results for the 
parameter determination at ILC$_{500}$, LHC and ILC$_{650}$.
We estimate errors in the mass measurements and
include statistical errors for the cross sections.
Fit programs, which exist so far only for the MSSM
\cite{fittino}, are not used.
Special attention is given to the interplay in combined analyses of the
experimental results at ILC and LHC.
 
\section{Strategy}
\label{sec:strategy}
We consider an NMSSM scenario where in the
gaugino/higgsino sector only the 
\linebreak
masses  
of the light chargino $\tilde{\chi}^{\pm}_1$  and
the light neutralinos
$\tilde{\chi}_1^0$ and
$\tilde{\chi}_2^0$
are accessible at the ILC$_{500}$.
We calculate the masses of the charginos and neutralinos and
the cross sections for the pair production 
of the light chargino $e^+e^- \rightarrow \tilde{\chi}^+_1
\tilde{\chi}^-_1$ 
and for 
the associated production of the light neutralinos
$e^+e^- \rightarrow \tilde{\chi}_1^0
\tilde{\chi}_2^0$ with polarized and unpolarized beams.

The masses and cross sections provide the experimental input
for deriving the supersymmetric parameters within the MSSM
using standard methods
\cite{choi,ckmz}:
\begin{itemize}
\item We assume an uncertainty of $1$--$2\%$ for the masses
$m_{\tilde{\chi}^{\pm}_1}$,
$m_{\tilde{\chi}^0_1}$, $m_{\tilde{\chi}^0_2}$,
$m_{\tilde{\nu}_e}$, $m_{\tilde{e}_L}$ and $m_{\tilde{e}_R}$.
In the cross sections for
$e^+e^- \rightarrow \tilde{\chi}^+_1 \tilde{\chi}^-_1$ and
$e^+e^- \rightarrow \tilde{\chi}_1^0 \tilde{\chi}_2^0$
we include
a statistical error of one standard deviation 
and take into account the error due to
the polarization uncertainty of $\Delta {\cal P}_{e^{\pm}}/{\cal P}_{e^{\pm}}=0.5\%$.

\item
From the chargino mass $m_{\tilde{\chi}^{\pm}_1}$ and 
the cross section 
$e^+e^- \rightarrow \tilde{\chi}^+_1 \tilde{\chi}^-_1$
measured at two energies, $\sqrt{s}=400$~GeV and $500$~GeV,
we determine bounds for the
elements $U_{11}$ and $V_{11}$ of the chargino mixing matrices.
Polarized beams allow the resolution of ambiguities and the improvement
of the accuracy. 
\item Using the mixing matrix elements $U_{11}$ and $V_{11}$,
the masses $m_{\tilde{\chi}^{\pm}_1}$, 
$m_{\tilde{\chi}^0_1}$ and $m_{\tilde{\chi}^0_2}$, 
and
the cross sections for
$e^+e^- \rightarrow \tilde{\chi}^0_1 \tilde{\chi}^0_2$, 
we derive constraints for
the parameters $M_1$, $M_2$, $\mu$ and $\tan\beta$. 
\item After the determination of the fundamental MSSM
parameters we calculate the heavy chargino
and neutralino masses and compare them with  
LHC analyses \cite{lhclc}.
\end{itemize}
The experimental results 
from the ILC with $\sqrt{s}=400$~GeV and $500$~GeV lead 
to consistent MSSM parameters despite the fact that we start with an NMSSM
scenario
where the light neutralino
$\tilde{\chi}^0_2$ has a significant singlino component.
In the considered scenario the ILC$_{500}$ does 
not allow a discrimination between MSSM and NMSSM
by mass and cross section measurements.
Then the identification of the underlying supersymmetric model
is performed 
in combined analyses at the LHC or the 
ILC$_{650}^{{\cal L}=1/3}$
by the measurement of the masses of the heavy neutralinos and
charginos. 

\section{Scenario}
\label{sec:scen}
We discuss an NMSSM scenario
where the lightest visible neutralino 
$\tilde{\chi}^0_2$ is accessible at the ILC$_{500}$ and has a significant
singlino component.
The supersymmetric parameters and the
resulting masses and neutralino mixing states are given in
Table~\ref{tab_nmssm_mssm}.
For the basis in the neutralino system we use the conventions of
\cite{Ellwanger:2004xm,Skands:2003cj}.
The hierarchy $M_1 > M_2$ of the U(1) and SU(2) mass parameters
causes an approximate mass degeneration
of the lightest neutralino $\tilde{\chi}_1^0$, which is assumed to be the
lightest supersymmetric particle (LSP), and the light chargino
$\tilde{\chi}_1^\pm$ that 
is typical for a minimal anomaly mediated SUSY-breaking mechanism (mAMSB)
\cite{AMSB}. 
Even small mass differences may be resolved experimentally with good accuracy
at the LHC \cite{amsb-lhc} and
by applying the ISR method \cite{Hensel} at the linear collider
\cite{amsb-lc}.
One should note, however, that we do not use any assumptions on 
possible SUSY-breaking mechanisms for the following parameter determination.
For comparison Table~\ref{tab_nmssm_mssm} also contains the parameters
of the 
MSSM scenario derived in Section 4, which  
leads to indistinguishable mass spectra and 
cross sections 
within the experimental errors at the ILC$_{500}$. 
All parameters are in agreement with bounds from 
dark matter constraints \cite{belanger}.

Table~\ref{tab_Higgs} shows the masses and mixings of the NMSSM Higgs
bosons for parameters $A_\lambda$ and $A_\kappa$ within the theoretical
and experimental constraints in the NMSSM scenario
\begin{eqnarray}
2740~\textrm{GeV} & < \; A_\lambda \; < & 5465~\textrm{GeV}, \\
-553~\textrm{GeV} & < \; A_\kappa \; < & 0~\textrm{GeV}.
\end{eqnarray}
Here the light scalar $S_1$ has MSSM-like character
while the second lightest scalar $S_2$ and the light pseudoscalar $P_1$
are almost pure singlets with masses larger than 200~GeV for
$-443~\textrm{GeV} < A_\kappa < -91~\textrm{GeV}$.
Within this $A_\kappa$ region 
direct Higgs production does not allow the identification of the 
NMSSM \cite{NMSSMhiggs} since
scalar and pseudoscalar Higgs bosons with dominant singlet character
escape detection. Detailed studies on the detection 
of light singlet dominated Higgs bosons at the LHC and
ILC can be found  in \cite{Miller:2004uh},
and of light pseudoscalars in scenarios with dominant
decay $S_1 \to P_1 P_1$ at LHC, ILC and the photon collider 
in \cite{S1toP1P1papers}.

Due to the mass difference between the light neutralinos in our
scenario the decays of  
$\tilde{\chi}^0_2$ and $\tilde{\chi}^0_3$
into $S_2$ and $P_1$ are kinematically forbidden.
The 
branching ratio for the decay of $\tilde{\chi}^0_2$ into the 
lightest Higgs boson $S_1$
is enhanced 
by a factor two to about $20\%$ in the NMSSM scenario compared to the MSSM.
Therefore precise measurements of the neutralino decays into Higgs bosons
could provide first evidence for the underlying model that has to be
confirmed by the combined LHC and ILC analyses described in the
next sections.
\begin{table}
\renewcommand{\arraystretch}{1.3}
\centering
\begin{tabular}{|l|c|c|}
\hline
& NMSSM & MSSM\\\hline\hline
$M_1$ & 360 GeV & 375 GeV \\
$M_2$ & 147 GeV & 152 GeV\\
$\tan\beta$ & 10 & 8 \\
$\mu$ & -- & 360 GeV \\
$\lambda$ & 0.5 & --\\
$x$ & 915 GeV & --\\
$\kappa$ & 0.2 & --\\
\hline\hline
$m_{\tilde{\chi}^0_1}$ & 138 GeV & 138 GeV \\
$m_{\tilde{\chi}^0_2}$ & 337 GeV & 344 GeV\\
$m_{\tilde{\chi}^0_3}$ & 367 GeV & 366 GeV\\
$m_{\tilde{\chi}^0_4}$ & 468 GeV & 410 GeV\\
$m_{\tilde{\chi}^0_5}$ & 499 GeV & -- \\
\hline
$m_{\tilde{\chi}^{\pm}_1}$ & 139 GeV & 139 GeV\\
$m_{\tilde{\chi}^{\pm}_2}$ & 474 GeV & 383 GeV\\ 
\hline
$m_{\tilde{e}_L}$ & 240 GeV & 240 GeV \\
$m_{\tilde{e}_R}$ & 220 GeV & 220 GeV \\
$m_{\tilde{\nu}_e}$ & 226 GeV & 226 GeV \\
\hline\hline
$\tilde{\chi}^0_1$ mixing state &
($-$0.02, 0.97, $-$0.20, 0.09, $-$0.07) & (0.03, $-$0.96, 0.26, $-$0.13)\\
$\tilde{\chi}^0_2$ mixing state &
(0.62, 0.14, 0.25, $-$0.31, 0.65) & (0.72, 0.22, 0.48, $-$0.46) \\
$\tilde{\chi}^0_3$ mixing state &
($-$0.75, 0.04, 0.006, $-$0.12, 0.65) & ($-$0.04, 0.10, $-$0.70, $-$0.71)\\
$\tilde{\chi}^0_4$ mixing state &
($-$0.03, 0.08, 0.70, 0.70, 0.08) & ($-$0.70, 0.18, 0.47, $-$0.52) \\
$\tilde{\chi}^0_5$ mixing state &
(0.21, $-$0.16, $-$0.64, 0.62, 0.37) & -- \\
\hline
\end{tabular}
\caption{Parameters, mass eigenvalues and neutralino mixing states
in the basis $(\tilde{B}^0, \tilde{W}^0, \tilde{H}^0_1,
\tilde{H}^0_2, \tilde{S})$
in the NMSSM scenario. 
For comparison the table also contains an MSSM scenario 
(neutralino basis $(\tilde{B}^0, \tilde{W}^0, \tilde{H}^0_1,
\tilde{H}^0_2)$)
with a similar mass spectrum of the
light neutralinos and charginos but different neutralino mixing.
\label{tab_nmssm_mssm}}
\end{table}

\begin{table}
\renewcommand{\arraystretch}{1.3}
\centering
\begin{tabular}{|l|c|}
\hline
$A_\lambda$ & 4000 GeV \\
$A_\kappa$ & $-$200 GeV \\
\hline
$m_{S_1}$ & 124 GeV \\
$m_{S_2}$ & 311 GeV \\
$m_{S_3}$ & 4347 GeV \\
$m_{P_1}$ & 335 GeV \\
$m_{P_2}$ & 4347 GeV \\
$m_{H^\pm}$ & 4346 GeV \\
\hline
$S_1$ mixing state & ($-$0.9947, $-$0.0990, 0.0292) \\
$S_2$ mixing state & ($-$0.0271, $-$0.0222, $-$0.9994) \\
$S_3$ mixing state & (0.0996, $-$0.9948,  0.0193) \\
$P_1$ mixing state & (0.0016,  0.0164, $-$0.9999) \\
$P_2$ mixing state & (0.0995,  0.9949,  0.0165) \\
\hline
\end{tabular}
\caption{The parameters $A_\lambda$ and $A_\kappa$ of the NMSSM Higgs sector
  and the resulting masses and mixing states of the
  Higgs particles according to NMHDECAY \cite{Ellwanger:2004xm} for the
  parameters in the neutralino sector as given in Table~\ref{tab_nmssm_mssm}
  and $M_Q = M_U = M_D = 800$~GeV and $A_t = A_b = 1500$~GeV.
\label{tab_Higgs}}
\end{table}

\section{Parameter determination at the Linear Collider}

In order to analyse the possible parameter reconstruction at the
ILC$_{500}$, we start with the NMSSM scenario of
Table~\ref{tab_nmssm_mssm} and calculate the resulting 
cross sections for neutralino and chargino production
for unpolarized beams and beam polarizations of
$({\cal P}_{e^-},{\cal P}_{e^+})=(- 90\%, +60\%)$ and
$(+ 90\%, -60\%)$ and assume 
$\Delta {\cal P}_{e^-}/{\cal P}_{e^-}=\Delta {\cal P}_{e^+}/{\cal P}_{e^+}=0.5\%$ \cite{Power}. 
At
$\sqrt{s}=400$~GeV chargino pairs
$\tilde{\chi}^{\pm}_1\tilde{\chi}^{\mp}_1$ are
produced, at $\sqrt{s}=500$~GeV the pairs
$\tilde{\chi}^{\pm}_1\tilde{\chi}^{\mp}_1$ and
$\tilde{\chi}^{0}_1\tilde{\chi}^{0}_2$ are visible.
Now the well-known strategies for determining the fundamental
MSSM parameters in the gaugino/higgsino sector
\cite{ckmz} are applied,
taking into account experimental
errors of 1.5\% (2\%) in the mass measurements of
$\tilde{\chi}^0_{2,3}$, $\tilde{e}_{L,R}$,
$\tilde{\nu}$ ($\tilde{\chi}^0_1$, $\tilde{\chi}^{\pm}_1$) and one 
standard deviation statistical errors for the cross sections
\cite{lhclc} as given in Table~\ref{tab_masscross}.
The statistical errors are based on a total luminosity of $500$~fb$^{-1}$,
which results in $\int {\cal L}=100$~fb$^{-1}$ for each of the 
possible five
polarization configurations $({\cal P}_{e^-},{\cal P}_{e^+})=(0, 0)$,
$(-90\%, +60\%)$, $(-90\%,-60\%)$, $(+90\%, +60\%)$ and $(+90\%,-60\%)$.

\begin{table}[t!]
\renewcommand{\arraystretch}{1.3}
\centering
\begin{tabular}{|l|c|l|c|}
\hline
$m_{\tilde{\chi}^0_1}$/GeV & $138\pm 2.8$ & $m_{\tilde{e}_L}$/GeV & 
  $240\pm 3.6$\\
$m_{\tilde{\chi}^0_2}$/GeV & $337\pm 5.1$ &$m_{\tilde{e}_R}$/GeV & 
  $220\pm 3.3$ \\
$m_{\tilde{\chi}^{\pm}_1}$/GeV & $139\pm 2.8$ & $m_{\tilde{\nu}_e}$/GeV & 
  $226\pm 3.4$\\
\hline\hline
\multicolumn{4}{|c|}{
$\sigma(e^+e^-\to\tilde{\chi}^{\pm}_1\tilde{\chi}^{\mp}_1)$/fb at 
 $\sqrt{s}=400$~GeV} \\ \hline
 \multicolumn{3}{|l|}{
Unpolarized beams} & $323.9 \pm 33.5$ \\
\multicolumn{3}{|l|}{
$({\cal P}_{e^-},{\cal P}_{e^+})=(-90\%, +60\%)$} & $984.0 \pm 101.6$  \\
\multicolumn{3}{|l|}{
$({\cal P}_{e^-},{\cal P}_{e^+})=(+90\%, -60\%)$} & $13.6 \pm 1.6$ \\
\hline \hline
\multicolumn{4}{|c|}{
$\sigma(e^+e^-\to\tilde{\chi}^{\pm}_1\tilde{\chi}^{\mp}_1)$/fb at 
 $\sqrt{s}=500$~GeV}\\ \hline
 \multicolumn{3}{|l|}{
Unpolarized beams} & $287.5 \pm 16.5$ \\
 \multicolumn{3}{|l|}{
$({\cal P}_{e^-},{\cal P}_{e^+})=(-90\%,+60\%)$} & $873.9 \pm 50.1$\\
 \multicolumn{3}{|l|}{
$({\cal P}_{e^-},{\cal P}_{e^+})=(+90\%,-60\%)$} & $11.7 \pm 1.2$\\
\hline \hline
\multicolumn{4}{|c|}{
$\sigma(e^+e^-\to\tilde{\chi}^{0}_1\tilde{\chi}^{0}_2)$/fb at
$\sqrt{s}=500$~GeV}\\ \hline
\multicolumn{3}{|l|}{ Unpolarized beams} & $4.0 \pm 1.2$ \\
\multicolumn{3}{|l|}{$({\cal P}_{e^-},{\cal P}_{e^+})=(-90\%,+60\%)$} & $12.1 \pm 3.8$ \\
\multicolumn{3}{|l|}{$({\cal P}_{e^-},{\cal P}_{e^+})=(+90\%,-60\%)$} & $0.2 \pm 0.1$ \\ \hline
\end{tabular}
\caption{Masses with 1.5\%
($\tilde{\chi}^0_{2,3}$, $\tilde{e}_{L,R}$, $\tilde{\nu}$)
and 2\% ($\tilde{\chi}^0_1$, $\tilde{\chi}^{\pm}_1$)
uncertainty and cross sections with an error
composed of
the error due to the mass and polarization uncertainties and
one standard deviation statistical error based on
$\int {\cal L}=100$~fb$^{-1}$, 
for both unpolarized beams and polarized beams with
$({\cal P}_{e^-},{\cal P}_{e^+})=(\mp 90\%,\pm 60\%)$ 
and $\Delta {\cal P}_{e^\pm}/{\cal P}_{e^{\pm}}=0.5\%$, 
in analogy to the study in \cite{lhclc}.
\label{tab_masscross}
}
\end{table}

\begin{figure}
\begin{center}
\setlength{\unitlength}{1cm}
\epsfig{file=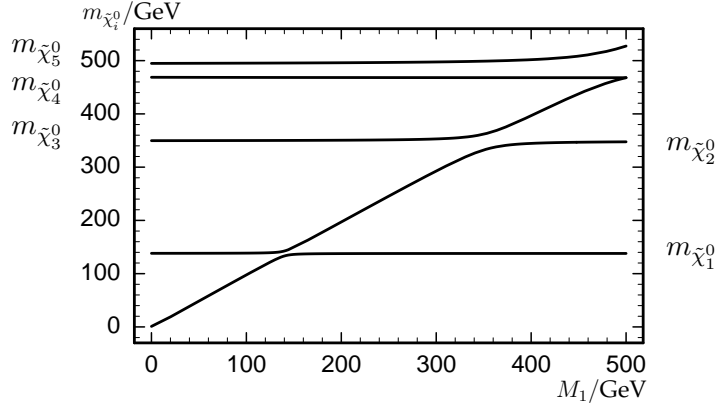,scale=.75}
\put(.2,2){\small $m_{\tilde{\chi}^0_1}$}
\put(.2,3.4){\small $m_{\tilde{\chi}^0_2}$}
\put(-8.5,3.6){\small $m_{\tilde{\chi}^0_3}$}
\put(-8.5,4.2){\small $m_{\tilde{\chi}^0_4}$}
\put(-8.5,4.7){\small $m_{\tilde{\chi}^0_5}$}
\end{center}
\caption{The neutralino masses $m_{\tilde{\chi}^0_i}$, $i=1,\ldots,5$
as a function of $M_1$ in the NMSSM for the scenario of
Table~\ref{tab_nmssm_mssm}.
\label{fig_m1} }
\end{figure}

We determine the components of the chargino eigenstates $U_{11}$ and $V_{11}$
\cite{Haber-Kane} from the measured polarized cross sections, which 
are bilinear quadratic functions of $U^2_{11}$ and $V^2_{11}$ and depend also
on  $m_{\tilde{\chi}^{\pm}_1}$ and $m_{\tilde{\nu}_e}$. 
The expressions for the chargino cross sections are analytically inverted
\cite{lhclc}.
In order to resolve ambiguities we assume that the process 
has been measured at two
different energies, $\sqrt{s}=400$ and $500$~GeV, and determine the
components of the chargino eigenstates, which are consistent with the
measured cross sections within the estimated error bounds of 
Table~\ref{tab_masscross}
\beqn
U^2_{11}=[0.84,1.0], &\quad& V^2_{11}=[0.83,1.0].
\label{eq_parnmssm}
\eeqn
We then exploit additionally the polarized neutralino cross sections
$\sigma(e^+e^-\to \tilde{\chi}^0_1 \tilde{\chi}^0_2)$ at
$\sqrt{s}=500$~GeV and the measured masses $m_{\tilde{\chi}^0_1}$,
$m_{\tilde{\chi}^0_2}$, $m_{\tilde{e}_L}$ and $m_{\tilde{e}_R}$ in order
to determine the parameters $M_1$, $M_2$, $\mu$ and $\tan\beta$:
\beqn 
M_1 &=& 377 \pm 42 \mbox{ GeV} \label{MSSMresult1},\\ 
M_2 &=& 150\pm 20 \mbox{ GeV},\\
\mu &=& 450 \pm 100 \mbox{ GeV},\\ 
\tan\beta &=& [1,30]. \label{MSSMresult4}
\eeqn
Note that, in our mAMSB-inspired scenario with $M_1 > M_2$, the crucial
observable to determine the parameter $M_1$ is $m_{\tilde{\chi}^0_2}$,
which can be clearly seen in Fig.~\ref{fig_m1} \cite{m1-paper}.
Since the heavier neutralino and chargino states are not produced,
the parameters
$\mu$ and $\tan\beta$ are determined with a considerable uncertainty.

In conclusion 
the measured masses and cross sections 
as experimental input from the ILC$_{500}$ 
are compatible with the MSSM. 
For the allowed ranges 
(\ref{MSSMresult1})--(\ref{MSSMresult4})
of the MSSM parameters we now predict the
heavier neutralino and chargino masses within the MSSM:
\beqn
m_{\tilde{\chi}^0_3} &=& 443\pm 107 \mbox{ GeV}, \label{eq_mp1_nm}\\
m_{\tilde{\chi}^0_4} &=& 490\pm 110 \mbox{ GeV},\\
m_{\tilde{\chi}^{\pm}_2} &=& 475 \pm 125 \mbox{ GeV}. \label{eq_mp3_nm}
\eeqn

We emphasize that although we started with an NMSSM scenario
where $\tilde{\chi}^0_2$ and $\tilde{\chi}^0_3$ have large
singlino admixtures (see Table~\ref{tab_nmssm_mssm}),
the strategy \cite{choi,ckmz} to determine the
MSSM parameters is successful.
All predictions for 
the heavier gaugino/higgsino masses are 
consistent in both models.
In order to distinguish between the models, the measurement of the 
heavier neutralino and chargino masses at the LHC or with  
the ILC$_{650}$ option is necessary.
In our NMSSM scenario the gaugino admixture of
$\tilde{\chi}^0_3$ probably allows 
its mass measurement at the LHC,
whereas in the MSSM the neutralino $\tilde{\chi}^0_3$
always has a strong higgsino character
within the predicted parameter ranges,
eqs.~(\ref{MSSMresult1})--(\ref{MSSMresult4}),
and would hardly be visible at the LHC.

\section{SUSY searches at the LHC}
\label{sec:lhc}
The LHC has an excellent potential to discover supersymmetric
particles.  The expected large cross sections for squark and gluino
production at the LHC
give access to a large spectrum of coloured as
well as non-coloured supersymmetric particles via the cascade decays.
Simulations \cite{sugra-lhc} exist e.g.\ for the mSUGRA point SPS1a
\cite{SPS}.
It has also been worked out how to measure the masses of the heavier
gauginos in
cascade decays at the LHC \cite{Giacomo}.  Mass
predictions from the ILC analysis lead to an
increase of statistical sensitivity for the LHC analysis and open the
possibility of identifying heavier gauginos even
in marginal signals in the squark cascades
\cite{lhclc}.  However, since higgsino-like charginos and neutralinos
do not couple to squarks,
their detection via cascade decays is not possible. Direct pair 
production of charginos and neutralinos 
that occurs via Drell--Yan production is strongly suppressed 
for masses larger than about 300 GeV as well as single neutralino 
production \cite{Gounaris:2004cv}.

The mAMSB-inspired scenario of 
Table~\ref{tab_nmssm_mssm}
contains nearly mass degenerate
$\tilde{\chi}^0_1$ and $\tilde{\chi}^{\pm}_1$
with an LSP--NLSP mass difference in the 
difficult
region for hadron colliders of
$200~\textrm{MeV} < m_{\tilde{\chi}^{\pm}_1} - m_{\tilde{\chi}^0_1}
< 2~\textrm{GeV}$
because the short-lived $\tilde{\chi}^{\pm}_1$ is not
directly observable and its SM decay particles are softly emitted
and suffer from a large background \cite{amsb-lhc}.
Applying specific cuts and using suitable kinematical variables, 
however, leads to a rather accurate measurement of the
mass difference. A gain in accuracy
can be obtained using the
LSP and NLSP mass from ILC measurements.
Since there exists no simulation
for our chosen parameter point, we assume 
a precision 
for the heavier neutralino masses 
as in \cite{Giacomo}, motivated by supersymmetric searches in AMSB scenarios
at the LHC \cite{amsb-lhc}.  
A more detailed analysis will be performed in a forthcoming work
\cite{Giacomo_forthcoming}.

The neutralinos $\tilde{\chi}^0_2$ and
$\tilde{\chi}^0_3$ have a large
bino-admixture and therefore appear
in the squark decay cascades. The dominant decay mode of
$\tilde{\chi}^0_2$ has a branching ratio $BR(\tilde{\chi}^0_2\to
\tilde{\chi}^{\pm}_1 W^{\mp})\sim 50\%$, while
for the $\tilde{\chi}^0_3$ decays $BR(\tilde{\chi}^0_3\to
\tilde{\ell}^{\pm}_{L,R} \ell^{\mp})\sim 45\%$ is largest.
Since the heavier neutralinos, $\tilde{\chi}^0_4$, $\tilde{\chi}^0_5$, 
are mainly higgsino-like,
no visible edges from these particles occur in the cascades.
It is expected to see the edges for
$\tilde{\chi}^0_2 \to \tilde{\ell}^{\pm}_R \ell^{\mp}$,
$\tilde{\chi}^0_2 \to \tilde{\ell}^{\pm}_L \ell^{\mp}$,
$\tilde{\chi}^0_3 \to \tilde{\ell}^{\pm}_R \ell^{\mp}$
and for
$\tilde{\chi}^0_3 \to \tilde{\ell}^{\pm}_L \ell^{\mp}$
\cite{Giacomo_forthcoming,Giacomo_private}.

With a precise
mass measurement of 
$\tilde{\chi}^0_1$,$\tilde{\chi}^0_2$, $\tilde{\ell}_{L,R}$ and
$\tilde{\nu}$ from the ILC$_{500}$ 
analysis, a
clear identification and separation of 
the edges of the two gauginos at the LHC is
possible without imposing specific model assumptions.
We therefore assume a precision of
about 2\% for the measurement of $m_{\tilde{\chi}^0_3}$, in analogy to
\cite{Giacomo}:
\beqn \label{eq:mchi3LHC}
&& m_{\tilde{\chi}^0_3}=
367\pm 7 \mbox{ GeV}.  
\eeqn 
Here the error assumption is rather conservative since
contrary to the scenarios in \cite{Giacomo} the
branching ratio
$BR(\tilde{q}_R \to \tilde{\chi}^0_3 q) \times
 BR(\tilde{\chi}^0_3 \to \tilde{\ell}^{\pm}_{L,R} \ell^{\mp})$
is very large
\cite{Giacomo_private}.

The precise mass measurement of $\tilde{\chi}^0_3$ is compatible with
the mass predictions 
of the ILC$_{500}$.
However, it is still to confirm that the measured particle is indeed the
$\tilde{\chi}_3^0$ and not the $\tilde{\chi}_4^0$.
The predicted $\tilde{\chi}_3^0$ in the MSSM is nearly a pure
higgsino, which is typical for the constrained MSSM, and
does not couple in the cascade decays, while the
$\tilde{\chi}_4^0$ with a sufficiently large gaugino component could be
measured in the cascades \cite{lhclc,Giacomo}. In order to identify
the particle and finally determine the masses and mixing characters of
the heavy neutralinos one has to discuss the following cases:
\begin{itemize}
\item Assuming the measured particle to be
$\tilde{\chi}^0_3$ and feeding it back in the ILC analysis lead to
improved parameter determination and mass predictions for
$m_{\tilde{\chi}^0_4}$ and $m_{\tilde{\chi}^{\pm}_2}$.
Using eq.~(\ref{eq:mchi3LHC}) for the ILC$_{500}$ analysis 
leads in our case, after rechecking with the allowed cross sections of
$\tilde{\chi}^0_1\tilde{\chi}^0_2$ and
$\tilde{\chi}^{+}_1\tilde{\chi}^-_1$ production, 
to the precise mass predictions:
\begin{equation}
m_{\tilde{\chi}^0_4}=[384, 393]\mbox{ GeV and }
m_{\tilde{\chi}^{\pm}_2}=[360, 380]\mbox{ GeV}.
\label{eq:heavymass-pred}
\end{equation} 
\item Assuming the measured particle to be
$\tilde{\chi}^0_4$ 
and feeding it back in the parameter determination of the ILC analysis
lead to inconsistency with the measured cross sections of
$\tilde{\chi}^0_1\tilde{\chi}^0_2$ and
$\tilde{\chi}^{+}_1\tilde{\chi}^-_1$ production.
\end{itemize}

The combined LHC/ILC$_{500}$ analysis therefore leads
to a consistent interpretation of the measured particles in the cascades.
However, a neutralino $\tilde{\chi}^0_3$  with sufficiently large
gaugino admixture to couple to squarks is incompatible with the
allowed parameter ranges of 
eqs.~(\ref{MSSMresult1})--(\ref{MSSMresult4}) in the MSSM.
One should note that this result has been derived within the
chargino/neu\-tralino sector of the CP-conserving unconstrained MSSM,
without any assumption concerning the SUSY-breaking mechanism.

We point out
that a measurement of the neutralino masses
$m_{\tilde{\chi}^0_1}$, $m_{\tilde{\chi}^0_2}$, $m_{\tilde{\chi}^0_3}$
which can take place at the LHC alone is not sufficient to distinguish
the SUSY models since rather similar mass spectra may exist.
Both a model-independent 
analysis using cross section and masses at the ILC and the 
precisely measured masses from the cascade analysis at the LHC are needed 
for the identification of the particles. 
Furthermore, the results from the LHC and the ILC$_{500}$ analyses
and the precise predictions for the missing 
chargino/neutralino masses, eq.~(\ref{eq:heavymass-pred}),
motivates to apply immediately the low-luminosity but higher-energy option
ILC$_{650}^{{\cal L}=1/3}$, which finally could lead to the correct
identification of the underlying model.

\section{The ILC$^{{\cal L}=1/3}_{650}$ option \label{LC650}}

The ILC design provides the option to enhance instantaneously the energy 
at the cost of
reduced luminosity. A centre-of-mass energy of
$\sqrt{s}=650$~GeV at a third of the luminosity,
i.e. $\int {\cal L}=33$~fb$^{-1}$ per polarization configuration,
does not require hardware
changes in the experimental set-up but only running time.  

The inconsistency of the MSSM analyses, 
the predicted higgsino-like $\tilde{\chi}^0_3$ and the measurement of
this neutralino
at the LHC, described in the previous sections
may motivate 
this higher-energy option.  The expected polarized and
unpolarized cross sections, including the statistical error on the basis of
one third of the luminosity of the ILC$_{500}$, are given in
Table~\ref{tab_heavy}.  
The neutralino
$\tilde{\chi}^0_3$ as well as the higgsino-like heavy neutralino
$\tilde{\chi}^0_4$ and the chargino $\tilde{\chi}^{\pm}_2$ are now 
accessible at the ILC$^{{\cal L}=1/3}_{650}$.
The production of
$\tilde{\chi}^0_1 \tilde{\chi}^0_3$ and 
$\tilde{\chi}^0_1\tilde{\chi}^0_4$ leads to 
promising rates at the 650~GeV option while the heaviest
neutralino $\tilde{\chi}^0_5$
is still too close to the threshold and
the production cross sections are probably too small to be observed.
The precise mass measurements of $\tilde{\chi}^0_4$ and
$\tilde{\chi}^{\pm}_2$
lead to a clear identification of the
supersymmetric model by comparing them with
the predictions of eq.~(\ref{eq:heavymass-pred}).
The accuracy of $\mu_{\rm eff}=\lambda x$ is improved
by  the measurements of the 
cross sections for the associated
$\tilde{\chi}^{\pm}_1\tilde{\chi}^{\mp}_2$ production
shown in Table~\ref{tab_heavychar}.  

Finally the 
masses $m_{\tilde{\chi}^0_1}$, $m_{\tilde{\chi}^0_2}$,
$m_{\tilde{\chi}^0_3}$, $m_{\tilde{\chi}^0_4}$ and
$m_{\tilde{\chi}^{\pm}_1}$, $m_{\tilde{\chi}^{\pm}_2}$, as well as the
cross sections, constitute the
observables for 
a fit of the 
parameters $M_1$, $M_2$, $\tan\beta$, $\lambda$
and $\mu_{\rm eff}=\lambda x$, $\kappa x$ 
of the gaugino/higgsino sector, which is, however, beyond the scope of
this paper.
Our analysis shows that the interplay between the two
experiments is crucial for the determination of the supersymmetric
model. It motivates
the use of the low-luminosity, $\sqrt{s}=650$~GeV, option of
the ILC in order to resolve model ambiguities even at an early stage of
the experiment and outlines future search strategies at an 
upgraded LHC as well as at the ILC at 1 TeV.

\begin{table}
\renewcommand{\arraystretch}{1.2}
\centering
\begin{tabular}{|c|c|c|c|}
\hline  &
 \multicolumn{3}{c|}{%
$\sigma(e^+e^- \to \tilde{\chi}^0_1\tilde{\chi}^0_j)/$fb at
$\sqrt{s}=650$~GeV}
 \\ \cline{2-4}
 & \makebox[21mm]{$j=3$} & \makebox[21mm]{$j=4$} & \makebox[21mm]{$j=5$} \\ 
\hline \hline 
Unpolarized beams & $12.2 \pm 0.6$
& $5.5 \pm 0.4$ & $\le 0.02$\\ \hline 
$({\cal P}_{e^-},{\cal P}_{e^+})=(-90\%,+60\%)$ &
$36.9 \pm 1.1$ & $14.8 \pm 0.7$ & $\le 0.07$\\ \hline 
$({\cal P}_{e^-},{\cal P}_{e^+})=(+90\%,-60\%)$ & 
$0.6 \pm 0.1$ & $2.2 \pm 0.3$ & $\le 0.01$\\ \hline
\end{tabular}
\caption{Expected cross sections for
the associated production of
the heavier neutralinos in the NMSSM  scenario of 
Table \ref{tab_nmssm_mssm} for the
ILC$^{{\cal L}=1/3}_{650}$ option with one sigma
statistical error 
based on ${\cal L} = 33$~fb$^{-1}$
for both unpolarized and polarized beams.
\label{tab_heavy}}
\end{table}

\begin{table}
\begin{center}
\renewcommand{\arraystretch}{1.2}
\begin{tabular}{|c|c|}
\hline
 & $\sigma(e^+e^- \to \tilde{\chi}^{\pm}_1\tilde{\chi}^{\mp}_2)/$fb at
$\sqrt{s}=650$~GeV\\
\hline\hline
Unpolarized beams & $2.4 \pm 0.3$ \\\hline
$({\cal P}_{e^-},{\cal P}_{e^+})=(-90\%,+60\%)$ & $5.8 \pm 0.4$ \\\hline
$({\cal P}_{e^-},{\cal P}_{e^+})=(+90\%,-60\%)$ & $1.6 \pm 0.2$ \\ \hline
\end{tabular}
\end{center}
\caption{Expected cross sections for the associated production
of the charginos
in the NMSSM scenario of 
Table \ref{tab_nmssm_mssm} 
for the ILC$^{{\cal L}=1/3}_{650}$ option
with one sigma statistical error 
based on ${\cal L} = 33$~fb$^{-1}$
for both unpolarized and polarized beams.
\label{tab_heavychar}}
\end{table}

\section{Conclusion}
\label{sec:con}
We have presented a scenario in the 
next-to-minimal supersymmetric standard model 
(NMSSM) that cannot be distinguished from the MSSM at
the first stage of the International Linear Collider with
$\sqrt{s}= 500$ GeV 
by cross section and mass measurements.
Although a light neutralino has a significant
singlino component in the NMSSM, the masses of the accessible
light neutralinos and charginos, as well as the production cross sections,
are identical in the two models within experimental errors.
The measurement of the masses of the heavier neutralinos and charginos
in combined analyses of the experimental results at the LHC and 
at the higher-energy option ILC$_{650}^{{\cal L}=1/3}$ with 
$\sqrt{s}= 650$~GeV leads to a clear identification of the supersymmetric
model.

\section*{Acknowledgements}

We are very grateful to G.~Polesello and P.~Richardson for
constructive discussions. We also thank J.~Kalinowski for
his critical reading of the manuscript.  
S.H.\ is supported by the G\"oran Gustafsson Foundation.
This work is supported by the
European Community's Human Potential Programme under contract
HPRN-CT-2000-00149, by the `Fonds zur F\"orderung der
wissenschaftlichen For\-schung' of Austria, FWF Project
No.~P16592-N02, and by the Deutsche Forschungsgemeinschaft (DFG) under
contract No.\ \mbox{FR~1064/5-2}.


\end{document}